# Continuous Electrical Manipulation of Magnetic Anisotropy and Spin Flopping in van der Waals Ferromagnetic Devices

Ming Tang[1,2†], Junwei Huang[1†], Feng Qin[1†], Kun Zhai[3], Toshiya Ideue[4,5*], Zeya Li[1], Fanhao Meng[1], Anmin Nie[3], Linglu Wu[1], Xiangyu Bi[1], Caorong Zhang[1,2], Ling Zhou[1], Peng Chen[1], Caiyu Qiu[1], Peizhe Tang[6,7], Haijun Zhang[2], Xiangang Wan[2], Lin Wang[3*], Zhongyuan Liu[3], Yongjun Tian[3], Yoshihiro Iwasa[4,8], Hongtao Yuan[1*]

[1]*National Laboratory of Solid State Microstructures, Jiangsu Key Laboratory of Artificial Functional Materials, College of Engineering and Applied Sciences, and Collaborative Innovation Center of Advanced Microstructures, Nanjing University, Nanjing 210000, China*

[2]*School of Physics, Nanjing University, Nanjing 210000, China.*

[3]*Center for High Pressure Science, State Key Laboratory of Metastable Materials Science and Technology, Yanshan University, Qinhuangdao 066000, China.*

[4]*Quantum Phase Electronic Center and Department of Applied Physics, The University of Tokyo, Tokyo 113-8656, Japan.*

[5]*Institute for Solid State Physics, The University of Tokyo, 5-1-5 Kashiwanoha, Kashiwa, Chiba 277-8581, Japan.*

[6]*School of Materials Science and Engineering, Beihang University, Beijing 100191, China.*

[7]*Max Planck Institute for the Structure and Dynamics of Matter, Center for Free Electron Laser Science, Hamburg 22761, Germany.*

[8]*RIKEN Center for Emergent Matter Science, Wako 351-0198, Japan.*

E-mail: htyuan@nju.edu.cn; ideue@ap.t.u-tokyo.ac.jp; linwang@ysu.edu.cn

[†] These authors contributed equally to this work.

**Controlling the magnetic anisotropy of ferromagnetic materials plays a key role in magnetic switching devices and spintronic applications. Examples of spin–orbit torque devices with different magnetic anisotropy geometries (in-plane or out-of-plane directions) have been demonstrated with novel magnetization switching mechanisms for extended device functionalities. Normally, the intrinsic magnetic anisotropy in ferromagnetic materials is unchanged within a fixed direction, and thus, it is difficult to realize multifunctionality devices. Therefore, continuous modulation of magnetic anisotropy in ferromagnetic materials is highly desired but remains challenging. Here, we demonstrate a gate-tunable magnetic anisotropy transition from out-of-plane to canted and finally to in-plane in layered $Fe_5GeTe_2$ by combining the measurements of the angle-dependent anomalous Hall effect and magneto-optical Kerr effect with quantitative Stoner–Wohlfarth analysis. The magnetic easy axis continuously rotates in a spin-flop pathway by gating or temperature modulation. Such observations offer a new avenue for exploring magnetization switching mechanisms and realizing new spintronic functionalities.**

The magnetic anisotropy of ferromagnetic (FM) materials, describing the magnetic moment along a preferential direction, plays a vital role in strongly correlated phenomena[1-3] and spintronic applications[4-6], such as magnetic tunneling junctions, spin–orbit torque, and spin–transfer torque devices. Examples of manipulating collinear spin orientation via a spin-flipping mechanism between spin up and down have been widely demonstrated in spintronic devices. Recently, a technical demonstration of three different types of spin orientations in spin–orbit torque devices[7] has inspired people to extend device functionalities within a single device. Based on this technical strategy, the electric field control of magnetic anisotropy can thus be a natural way to realize *in situ* continuous control of spin orientation and magnetization switching in a spintronic device. Essentially, such magnetic anisotropy can be significantly affected by the surrounding environment of magnetic sites, such as the local symmetry within the atomic monolayer[8,9], the specific electronic states at the Fermi level[10,11], or the exchange interactions between magnetic sites nearby[12,13]. Owing to the remarkable and sophisticated interplay between the itinerant electrons and localized magnetic moments, the itinerant electron cloud can surround and screen the local magnetic moment and can serve as a tunable physical parameter of ferromagnetism therein. Note that the strength of such screening is determined by the density of state (DOS) of itinerant electrons at the Fermi level[14-16], which can be modulated by an external electric field. Once the local environment for the screening effect changes in a certain way, the distribution of electron orbitals thus varies, resulting in a magnetic anisotropy transition between in-plane and out-of-plane FM states. Consequently, the preferred spin orientation (also the magnetic easy axis) will rotate continuously or discontinuously via spin-flopping or spin-flipping transition[17], depending on the evolution of the free energy minima of spin states in FM materials, as schematically shown in Fig. 1a. Based on such a scenario, magnetization switching with a tunable magnetic easy axis can serve as a powerful way to realize practical multifunctionality devices[7,18]. Tuning magnetic anisotropy in itinerant FM materials based on many-body interactions between itinerant and localized electrons is thus of great significance in physics and highly desired for practical device applications; however, it remains challenging.

Herein, we demonstrate gate-tunable magnetic anisotropy with continuous control of the magnetic easy axis between out-of-plane and in-plane in a layered FM material, Fe$_5$GeTe$_2$ (F5GT), by combining anomalous Hall effect (AHE) and magneto-optical Kerr effect (MOKE) measurements, as well as quantitative Stoner–Wohlfarth analysis of magnetic anisotropy energy (MAE). We find that the itinerant electron density in layered F5GT can be tuned to drive the transitions between magnetic states with different magnetic easy axes. In particular, the MAE can be modulated in a large range from 2.11 to −0.38 MJ/m$^3$, with $\theta_{EA}$ changing from 0° to 90° (where $\theta_{EA}$ is the angle of the magnetic easy axis relative to the normal direction of the sample surface). Interestingly, $\theta_{EA}$ has been confirmed to rotate continuously in a spin-flop pathway by tuning the screening effect from itinerant electrons through changing the DOS at the Fermi level. Such continuous control of $\theta_{EA}$ ensures the designable intermediate states of magnetic anisotropy and can further serve as a powerful magnetization switching mechanism. Our results promote the understanding of the underlying mechanism of the interplay between itinerant and localized electrons and the resulting controllable MAE in van der Waals FM materials, paving the way to the application of designing spintronic devices with new functionalities.

We chose layered FM F5GT as our studied system for the following three reasons. First, in the monolayer rhombohedral F5GT structure (Fig. 1b), the Fe atoms (occupying three inequivalent Fe1, Fe2, and Fe3 sites) and the Ge atoms are distributed in seven atomic layers, forming a covalently bonded $Fe_5Ge$ slab sandwiched between two Te layers (more details in Supplementary Figs. 1, 2). Such an Fe-rich structure generates complex magnetic orders, and the magnetic easy axis of its bulk form prefers the in-plane direction[19,20], providing a new platform to understand the MAE in FM systems. Second, the magnetic ground states of FM F5GT can be perturbed by the many-body interactions between the coexisting itinerant and localized $3d$ electrons. The tuning of the DOS of itinerant electrons at the Fermi level can modulate the screening effect on localized $3d$ electron moments[16] and is expected to affect the spin–orbit coupling and the resultant MAE. Third, the coexistence and competition of the charge order and ferromagnetism in F5GT have been confirmed at temperatures below 150 K, where a $\sqrt{3} \times \sqrt{3}$ superstructure appears and is attributed to Fermi surface nesting[21,22]. Thus, the interplay between the charge order and ferromagnetism in F5GT can be modulated by changing the DOS of itinerant electrons at the Fermi level, resulting in a prospective magnetic anisotropy transition.

To understand the fundamental magnetic anisotropy properties of F5GT, we systematically measured the AHE resistance $R_{AHE}$ by varying the temperature and magnetic field $\mu_0 H$. As shown in Fig. 1c, as the temperature decreases, the linearly shaped $R_{AHE}$–$\mu_0 H$ curve of a flake (86 nm) becomes nonlinear with a tiny hysteresis from 280 K (more details are provided in Supplementary Fig. 3). The Curie temperature ($T_C$) of 286 K is confirmed by the Arrott plot (Supplementary Fig. 4), consistent with previous reports[19,20]. The hysteresis loops display relatively tilted shapes above 150 K, implying an in-plane magnetic easy axis therein. In general, the shape of the hysteresis loop in an FM system is determined by the domain structures, resulting from the balance of magnetic anisotropy energy, dipolar energy, and exchange energy, in which stripe domain patterns are frequently observed[23]. To directly determine the orientation of the magnetic easy axis in F5GT, we performed angle-dependent AHE measurements and quantitative analysis of MAE with the phenomenological Stoner–Wohlfarth model. As described in the Methods section, this model can provide the following key information on the magnetic easy axis (more details in Supplementary Table 1) and the specific shape of AHE hysteresis loops (more details in Supplementary Figs. 5, 6) with the first-order constant $K_1$ and second-order constant $K_2$ of MAE: i) $K_1 > 0$ (or $K_1 < 0$) represents FM ordering with the magnetic easy axis oriented along the out-of-plane (or in-plane) direction; ii) $K_1$ together with $K_2$ can describe the FM ordering state with the canted angle $\theta_{EA}$ for the magnetic easy axis. The $K_1$ and $K_2$ values can be obtained from the relation between $\theta_H$ and $\theta_M$ through angle-dependent $R_{AHE}$–$\mu_0 H$ measurements, where $\theta_H$ (or $\theta_M$) is the angle between the direction of the magnetic field $\mu_0 H$ (or the magnetization $M$) and the normal direction of the sample surface. On this basis, we utilized the Stoner–Wohlfarth model to analyse the raw $R_{AHE}$–$\mu_0 H$ curves at different $\theta_H$ (Fig. 1d) so that $\theta_M$–$\theta_H$ can be obtained in Fig. 1e. Taking the results of the flake (86 nm) at 210 K as an example, the observation that the $\theta_M$–$\theta_H$ data fall into the region of $\theta_M > \theta_H$ directly indicates an in-plane magnetic easy axis, which is further verified by the fitted negative $K_1$ value of −0.202 MJ/m³. Such an in-plane magnetic easy axis observed in 86-nm-thick F5GT at 210 K is consistent with previous reports in bulk F5GT crystals with magnetization measurements[19,20].

To gain insight into the domain character of the F5GT flake with an in-plane magnetic easy axis, we obtained magnetic domain images and MOKE loops in a longitudinal geometry with a wide-field Kerr microscope. The in-plane magnetization and the resulting Kerr rotation in the MOKE measurements can be directly reflected by the intensity contrast of the magnetic domain images (more details in Supplementary Fig. 7). Figure 1f shows the magnetic domain images of an F5GT flake at 150 K, visualizing the evolution of the in-plane component of magnetization during the sweeping of an in-plane magnetic field. It is evident that there exist large magnetic domains on the scale of tens of micrometers from the magnetic domain images of F5GT (the domains in dark regions are labelled by the dashed yellow lines), excluding the dominance of domain walls and dipolar energy on the magnetic states. As shown from the top to bottom magnetic domain images (more details in Supplementary Fig. 8 and Supplementary Movie 1), the in-plane magnetic field was initially applied to 100 mT to fully magnetize the sample before sweeping the in-plane magnetic field. The magnetic domain image of the F5GT flake starts from the whole dark region, representing a single magnetic domain whose magnetization is parallel to the initial magnetic field. It gradually changes to mix with incremental bright areas (representing the mixed magnetic domains with different magnetization directions) and eventually reaches an entire bright region (representing the magnetic domain whose magnetization is completely parallel to the applied external magnetic field). To check the magnetization evolution with the magnetic field, we present the MOKE hysteresis loop in the bottom panel of Fig. 1f, in which the signal for the Kerr angle is obtained by integrating the intensities of the whole flake region in magnetic domain images while sweeping the in-plane magnetic field (more details in Supplementary Fig. 9). One can see that the MOKE loop represents a rectangular shape with a small value of the coercive magnetic field $\mu_0 H_c \sim 3$ mT. Such a rectangular shape in the MOKE loop, distinctive from the relatively tilted shape in the AHE loop with sweeping of the out-of-plane magnetic field (Fig. 1c), further confirms the in-plane magnetic easy axis in F5GT.

To understand the evolution of the in-plane magnetic anisotropy with temperature, we show magnetic domain images under different temperatures in Fig. 2a. As shown in the bottom panel of Fig. 2a, the magnetization and the resultant Kerr angle can be directly reflected by the intensity contrast of magnetic domain images (quantitatively reflected by the amplitude of the MOKE loops). The strongest intensity contrast is observed at 150 K and confirmed by the amplitude of MOKE loops, which can be attributed to the temperature-dependent evolution of magnetic anisotropy, as discussed below. To quantitatively understand the evolution of magnetic anisotropy with temperature, we performed angle-dependent AHE measurements at various temperatures and calculated the magnetic anisotropy constant $K_1$. As shown in Fig. 2b, with decreasing temperature, the $K_1$ values change from negative to positive at a temperature of approximately 150 K (denoted as $T_M$), serving as direct evidence of the magnetic anisotropy transition from in-plane to out-of-plane (more details in Supplementary Fig. 10). With further cooling down to the typical temperature at ~ 25 K (denoted as $T_K$), the $K_1$ values change back to negative values, indicating a magnetic anisotropy transition from out-of-plane back to in-plane, which might be related to the Kondo screening effect[24,25]. Specifically, in Fig. 2c, the saturated anomalous Hall resistance $R_{AHE}^s$ reaches the maximum around $T_M$ and drops to almost zero at lower temperatures, deviating from the classic saturation behavior of magnetization described by Bloch's law (more details in Supplementary Fig. 11). This deviation can be attributed to the change of the magnetic easy axis from in-plane to out-of-plane. Correspondingly, at $T_M$, the remnant anomalous Hall resistances $R_{AHE}^r$ show a sudden upturn from zero. Note that such remarkable changes in $R_{AHE}^s$ and $R_{AHE}^r$ occur simultaneously at $T_M$, consistent with the temperature where the sign of $K_1$ changes from negative to positive.

Since the 3$d$ electrons contributed by Fe atoms dominate the ferromagnetism in F5GT (ref.[22]), the DOS change of itinerant electrons at the Fermi level should strongly impact the states of 3$d$ electrons. This might potentially result in the magnetic anisotropy transition. Therefore, it is significant to investigate the evolution of the Hall coefficient $R_H$ to understand the relationship between the DOS of itinerant electrons and magnetic anisotropy. As shown in Fig. 2d, the $R_H$ values decrease monotonically during cooling (more details of deducing $R_H$ values are provided in Supplementary Figs. S12−14). Interestingly, near $T_M$, the positive $R_H$ becomes negative (more details in Supplementary Fig. 15), suggesting the appearance of new electron pockets associated with the emergence of charge order, as demonstrated in recent reports[21,22]. Such formation of new electron pockets might contribute to the drop in the temperature-dependent longitudinal resistance $R_{xx}$ near $T_M$, as shown in Fig. 2d. The consistency of these transitions at $T_M$ shown in Fig. 2, b–d provides direct evidence that the magnetic anisotropy transition is strongly correlated with the change in DOS. Specifically, when cooling the temperature to $T_M$, a new electron pocket (corresponding to the $d_{xz}$ and $d_{x^2-y^2}$ orbitals of Fe1 atoms[22]) appears at the M point and leads to the electron occupation of those atomic orbitals. As a result, the DOS of the intinerant electrons in such a system can be changed dramatically with temperature cooling down to $T_M$, which changes the contribution of the spin−orbital coupling to the magnetic moments and causes the magnetic anisotropy transitions between in-plane and out-of-plane magnetic phases. Additionally, as indicated by the black arrow, the $R_H$ values show an additional drop near $T_K$, indicating the emergence of hole pockets and a sudden change in the itinerant electron density. Such consistent behavior of the temperature-dependent evolution of $R_H$ and $K_1$ near $T_M$ and $T_K$ evidently provides an opportunity for modulating the magnetic anisotropy by tuning the DOS in F5GT. We argue that the change in the DOS modulates the screening strength on magnetic moments, which possibly drives magnetic anisotropy transitions between in-plane and out-of-plane.

To understand the effect of flake thickness on the magnetic anisotropy transition, we show the temperature-dependent $K_1$ values in Fig. 2e for flakes with different thicknesses. One can clearly see that the $K_1$–$T$ curves directly present the magnetic anisotropy transition between in-plane and out-of-plane in thick flakes (130 nm and 86 nm) and gradually show upward shifts in the whole temperature region when the flake thickness decreases to 18 nm. With reduced thickness, the critical temperature $T_M$ increases rapidly to higher temperatures, while the critical temperature $T_K$ decreases (Supplementary Fig. 16). Such an evolution of the $K_1$ values, as well as the change of $T_M$ and $T_K$, implies that the magnetic easy axis in F5GT can be effectively modulated by the flake thickness, similar to the results reported in the layered 1T-CrTe2 (ref. [26]). Correspondingly, the temperature-dependent $\theta_{EA}$ for the above flakes, which can quantitatively describe the direction of the magnetic easy axis, provides specific information on the thickness-dependent magnetic easy axis, as shown in the lower inset of Fig. 2e. Remarkably, $\theta_{EA}$ for the thick flake (86 nm) can have intermediate values between 90° and 0° around $T_M$ and $T_K$, suggesting the existence of a canted magnetic easy axis. $\theta_{EA}$ for the thin flake (18 nm) maintains a constant value of 0°, implying that the magnetic easy axis maintains the out-of-plane position for the whole temperature region. These observations suggest that thinner flakes prefer out-of-plane magnetic anisotropy, in sharp contrast to thicker flakes, which favor in-plane magnetic anisotropy. To further confirm such thickness-dependent magnetic anisotropy, we performed an alternative measurement of polar MOKE. One can see a remarkable thickness-dependent difference in the shape of hysteresis loops in F5GT in the upper inset of Fig. 2e. The hysteresis loop of the thick flake (89 nm) displays a line shape, which corresponds to the magnetic hard axis. In comparison, the hysteresis loop of the thin flake (23 nm) is rather rectangular, corresponding to the out-of-plane magnetic easy axis. Such observations are consistent with the electronic transport results in Fig. 2e.

To realize the applications of F5GT in magnetic switching and spintronics devices, we performed electrolyte-gating experiments to electrically control the magnetic anisotropy by tuning the DOS around the Fermi level. Figure 3a shows a photograph of an F5GT device with a flake thickness of 42 nm and a schematic cartoon of the electrolyte-gating geometry. A solution of LiClO$_4$ and polyethylene oxide (PEO) was used as the gating electrolyte. At 330 K, the lithium ions (Li$^+$) are expected to be electrically driven into the van der Waals gaps of bulk crystals without chemical reactions to the monolayers therein[27,28] and finally cause extreme electron doping in bulk form for tuning the magnetic properties of F5GT flakes. As shown in Fig. 3b, the $R_{\mathrm{AHE}}^{\mathrm{s}}$ values (orange balls and line) at 2 K first increase as the gate voltage $V_{\mathrm{G}}$ increases to 1.70 V and suddenly decrease when $V_{\mathrm{G}} > 2.20$ V. Such an effective tuning of $R_{\mathrm{AHE}}^{\mathrm{s}}$ indicates that the electric field indeed can modulate the magnetic states in F5GT, which should be attributed to the change in DOS of the itinerant electrons around the Fermi level. Note that the magnetic properties, which are detected by the Hall effect and magnetoresistance measurements, are not sensitive to the stacking faults[29] (more details in Supplementary Note 1). Therefore, the manipulation of magnetic anisotropy mainly originates from the tuning of carrier density in gated F5GT samples, as confirmed by our gate-dependent Hall coefficient results. We plot the Hall coefficient $R_{\mathrm{H}}$ mapping as a function of $V_{\mathrm{G}}$ and temperature in Fig. 3b. The sudden drop of $R_{\mathrm{H}}$ around $T_{\mathrm{M}}$ disappears when $V_{\mathrm{G}} > 2.20$ V (more details in Supplementary Fig. 17), which might be ascribed to the suppression of the charge order by the gate voltage. In addition, the isoline of a certain $R_H$ value (light blue color in the contour map) matches well with the $R_{AHE}^s$–$V_G$ relation, which confirms that the change in the magnetic state comes from the DOS change. Figure 3c shows the gate-tunable AHE hysteresis loops and magnetoresistance of F5GT at 2 K. At zero gate voltage, the shape of the AHE hysteresis loop is rectangular. With increasing $V_G$, such a rectangular hysteresis loop starts to tilt when $V_G \geq 2.00$ V and eventually becomes a titled line shape without any hysteresis at those $V_G$ higher than 2.50 V (left panel of Fig. 3c). This evolution indicates that the magnetic easy axis reorients from out-of-plane to in-plane. Correspondingly, the cusped butterfly characteristics of magnetoresistance start to be broader with $V_G \geq 2.00$ V and disappears at higher $V_G$ (right panel of Fig. 3c and Supplementary Fig. 18). The evolution of the AHE and magnetoresistance validates that the tuning of the DOS can successfully drive the transitions of magnetic states in F5GT at 2 K. Based on the well-known scaling theory of AHE, the relationship between the Hall and longitudinal conductivity of F5GT (the $\sigma_{xy}$–$\sigma_{xx}$ plot, more details in Supplementary Figs. 19, 20) is located near the boundary between the dirty metal region (originating from impurity scattering events) and the intrinsic region (attributed to intrinsic dissipationless topological Berry phase)[30-32], displaying a gate-tuned unconventional dome-shape behavior, which provides insight into the complicated origins of ferromagnetism in F5GT.

More importantly, to quantitatively understand the electrical tuning of the magnetic anisotropy, we analysed the temperature-dependent $K_1$ values under different $V_G$ values in a device with a thickness of 23 nm (Fig. 3d). One can see that the magnetic anisotropy is tuned from out-of-plane to in-plane by applying gate voltage at low temperatures (more details in Supplementary Fig. 21). Three important features need to be addressed here: i) at $V_G = 0$ V, the $K_1$ values remain positive in the whole temperature region; ii) as $V_G$ increases, the $K_1$ values decrease within the whole temperature range, implying that the magnetic anisotropy of F5GT can be dramatically tuned by gating; iii) as $V_G$ increases to above 2.48 V, the $K_1$ value at 2 K becomes negative, directly indicating that the magnetic easy axis is effectively tuned from the out-of-plane to the in-plane by gating. To understand the microscopic process of the gate-induced magnetic anisotropy transition, we calculate the magnetic anisotropy constants $K_1$ and $K_2$ at 2 K for different $V_G$ and further analyse the gate-dependent canted angle of the magnetic easy axis $\theta_{EA}$. As shown in the lower panel of Fig. 3d, $\theta_{EA}$ equals 0° (out-of-plane magnetic easy axis) when $V_G \leq 1.93$ V. As $V_G$ further increases to 2.55 V, $\theta_{EA}$ gradually rotates from 0° to 90°, indicating that the initial out-of-plane magnetic easy axis is tuned in-plane. Since the demagnetization energy and surface anisotropy energy (which are related to sample shape and thickness) do not change for this particular device (more details in Sepplementary Fig. 22), the magnetic crystalline anisotropy energy can be directly tuned by the gating effect. Such a continuous tuning of $K_1$ values opens up opportunities for understanding the interplay between itinerant electrons and localized $3d$ electrons[33,34] and sheds light on potential voltage-controlled magnetic devices.

Since the applied gate voltage can modulate the DOS of itinerant electrons around the Fermi level[28], the environment surrounding the localized magnetic moment will be altered accordingly. This will result in the change of distribution for those $3d$ orbitals around the Fermi level and lead to the modulation of the screening effect from itinerant electrons to those localized magnetic moments, both of which directly determine the magnetic anisotropy and the $\theta_{EA}$. Therefore, as shown in gate-dependent $R_H$ (Fig. 3b) and $\theta_{EA}$ (Fig. 3d), the consistency in the critical $V_G$ value at approximately 2 V reveals a strong correlation between the magnetic anisotropy and the itinerant electron DOS connected with the screening effect in F5GT. In general, the screening effect can be modulated by both the DOS of itinerant electrons and the flake thickness[28]. To provide more support for such a correlation, the similarities between the temperature-dependent $\theta_{EA}$ for different flake thicknesses and gate voltages are shown in Fig. 3e. Compared to a thinner flake (18 nm) where $\theta_{EA}$ stays at 0° (out-of-plane), $\theta_{EA}$ for a thicker flake (86 nm) evolves from 0° at approximately $T_K$ to 90° during cooling (upper panel of Fig. 3e). Interestingly, owing to the gate-enhanced DOS at the Fermi level, the itinerant electron screening effect dominates the MAE: the $\theta_{EA}$ of the pristine flake with a thickness of 23 nm remains at 0° throughout the tested temperature range ($V_G$ = 0 V), while rotation of the magnetic easy axis emerges at approximately $T_K$ at $V_G$ = 2.55 V (lower panel in Fig. 3e). Both the increased thickness and the applied $V_G$ enhance the screening effect in F5GT, resulting in a change of $\theta_{EA}$ from 0° to 90°. The fact that dimensions and gating have similar impacts on MAE suggests that the DOS and screening effect play essential roles in the MAE.

Figure 4a highlights the $V_G$-dependent evolution of $K_1$ and $K_2$, as well as the resulting spin-flop transition induced by the gate-controlled screening effect at 2 K. From the $K_1$–$K_2$ plots in Fig. 4a, one can see that there are three regions for different spin orientation states, namely, the in-plane (shaded in blue), out-of-plane (shaded in pink), and canted (shaded in yellow) spin orientations. The data points of $K_1$ = 2.26 MJ/m³ and $K_2$ = −0.747 MJ/m³ for F5GT (23 nm) at $V_G$ = 0 V and $T$ = 2 K (red balls) fall into the region with the out-of-plane magnetic easy axis. Starting from such an initial out-of-plane spin state, two different pathways for the magnetic transition toward the in-plane spin state are presented in Fig. 4a: the spin-flop transition with an intermediate canted spin state along the anticlockwise direction and the spin-flip transition along the clockwise direction. For gated F5GT, the spin state confirmed by $K_1$ and $K_2$ evolves anticlockwise with the applied $V_G$ along the spin-flop transition pathway. Notably, at temperatures higher than $T_K$ (blue balls at 50 K), the $K_1$ and $K_2$ values can only be tuned in a small range, and thus, there is no magnetic anisotropy transition at all. Such experimental observations provide unambiguous evidence of gate-controlled magnetic anisotropy in F5GT via the gradual evolution of the DOS of the itinerant electron at the Fermi level[28,35].

Based on the survey of the $K_1$ values of MAE for well-known FM materials (Fig. 4b and Supplementary Tables 2–4), we notice that the $K_1$ values in most FM materials remain positive with out-of-plane magnetization and usually become larger while reducing the flake thickness. Surprisingly, in iron-based FM materials Fe$_n$GeTe$_2$ ($n$ = 4, 5), the $K_1$ values can change between positive and negative, indicating that the magnetic easy axis can change between out-of-plane and in-plane. The MAE of magnetic materials is determined by the spin–orbit coupling in 3$d$ orbitals, which is sensitive to the DOS at the Fermi level and thus the filling factor of 3$d$ electrons on Fe atoms[33,36]. In these Fe$_n$GeTe$_2$ materials, the coupling or competition between itinerant electrons and localized 3$d$ electrons starts to be complicated. Such a dependence of magnetic anisotropy on the Fe stoichiometry, temperature, flake thickness[19,37,38], or gate voltage (this work) all results from the DOS change at the Fermi level. In particular, the MAE can be tuned due to massive modulation of the DOS of itinerant electrons at the Fermi level (Fig. 3b and Supplementary Fig. 21), indicating that the gate voltage tunes the hybridization strength between itinerant electrons and localized 3$d$ electrons. Notably, the giant $K_1$ values of the out-of-plane MAE of 2.83 MJ/m$^3$ in F5GT thin flakes can be electrically modulated in an extensive range of several MJ/m$^3$ (from −0.38 to 2.11 in the gating device), which is almost the largest value among the reported 3$d$ FM materials and close to the values of widely used $f$-orbital permanent magnets such as Nd$_2$Fe$_{14}$B (ref. [39]), paving a new way for practically tunable spintronic applications. In short, we demonstrate electrical engineering of the magnetic anisotropy transition from an out-of-plane to an in-plane magnetic easy axis via modulation of the itinerant electron screening effect in layered F5GT, where the $K_1$ values can be effectively tuned from positive to negative by gating. The magnetic properties and complicated evolution of $K_1$ with temperature, thickness, and gate voltage may indicate the complex magnetic states in F5GT beyond the description of the Stoner–Wohlfarth model for a single magnetic domain, in which the complex magnetic states are possibly caused by a noncollinear magnetic order[40]. Combining the near-room-temperature $T_C$ with the itinerant nature of strongly correlated electrons in such an FM material, the gate-tunable DOS of itinerant electrons at the Fermi level and the high tunability of MAE, our findings will shed light on the achievements of new electronic functionalities and device architectures for spin–orbit torque or spin–transfer torque devices, magnetic tunnel junctions, and nonvolatile memories[6,41-43].

## Methods

**Device fabrication of F5GT flakes.** High-quality F5GT single crystals were grown by the chemical vapour transport method. The crystal structure was characterized by atomic-resolution scanning transmission electron microscopy (STEM), as shown in Supplementary Fig. 1. Ti/Au (3/12 nm) Hall-bar electrodes were patterned on a $SiO_2$/Si substrate through standard electron-beam lithography and metal evaporation procedures. The F5GT flakes were exfoliated onto PDMS (polydimethylsiloxane) and transferred onto prepatterned substrates. Then, they were immediately capped with *h*-BN and spin-coated with PMMA (polymethyl methacrylate) for protection from air. All exfoliation and transfer processes were performed inside a $N_2$ gas glovebox (oxygen and water levels below 0.1 ppm). The thickness of the F5GT flakes was confirmed through atomic force microscopy (AFM).

**Electrolyte gating of F5GT flakes.** A solution of $LiCO_4$ and PEO was used as the gating electrolyte, where a mixed powder of 0.1 g $LiClO_4$ and 0.33 g PEO was dissolved in 5 ml anhydrous methanol. The Li ions can be electrically driven into the van der Waals gap, realizing extreme electron doping[27,28]. The electrolyte was stirred over 24 hours and kept at 50 °C before use. After dropping the electrolyte on the F5GT sample surface, the samples were kept in vacuum at 100 °C for 0.5 hours and annealed to remove the methanol solvent before loading into the vacuum chamber for transport measurements.

**Electronic transport measurements.** Electronic transport measurements were conducted in an Oxford Instruments TeslatronPT system equipped with a 12 T superconducting magnet. During our magnetoresistance and Hall effect resistance measurements for pristine F5GT, the devices were directly zero-field cooled from 300

K to the base temperature of 1.5 K with a cooling rate of 1 K/min. All temperature-dependent magnetic measurements were performed at specific temperatures during the warm-up process. For gated F5GT, the devices were warmed up to 370 K to in situ annealing for 30 minutes to remove the organic solvent of the electrolyte, and the gate voltage was applied at 330 K before cooling down to 1.5 K, followed by magnetic measurements during the warm-up process. The angle-dependent anomalous Hall effect (AHE) measurement was carried out with a rotating sample stage with a high angle resolution better than 0.01°. Synchronized lock-in amplifiers (SR830, Stanford Research Systems) were used to detect the longitudinal and Hall voltages. A DC source meter (Keithley 2400) was used to supply the gate voltage and measure the leakage current simultaneously. The AHE is routinely employed to understand the magnetic properties of small flakes in the Hall-bar geometry. Generally, the acquired Hall resistance $R_{xy} = V_y/I_x$ can be written as $R_{xy} = R_H \mu_0 H + R_{AHE}$, where $R_H$ is the Hall coefficient and $R_{AHE}$ is the anomalous Hall resistance.

The Hall resistance $R_{xy}(H)$ is obtained from the antisymmetrized component of the raw data of Hall resistance $R_{raw}(H)$ with the following procedure: $R_{xy}(H) = [R_{raw}(+H) - R_{raw}(-H)]/2$ and $R_{xy}(-H) = [R_{raw}(-H) - R_{raw}(+H)]/2$, in which the raw data of Hall resistance $R_{raw}(+H)$ is the half loop sweeping from the positive field to the negative field, and $R_{raw}(-H)$ is the half loop sweeping from the negative field to the positive field. As a result, the Hall resistance $R_{xy}(H)$ is antisymmetric with the magnetic field satisfying $R_{xy}(+H) = -R_{xy}(-H)$. Based on the relation $R_{AHE} \propto (R_S M)_\perp$, $R_{AHE}$ can directly reflect the out-of-plane component of magnetization, where $R_S$ and $M$ are the coefficients characterizing the strength of $R_{AHE}$ and the magnetization, respectively. The normal Hall resistance $R_H \mu_0 H$ contributes to the linear background to $R_{xy}(H)$, and the slope corresponds to the Hall coefficient $R_H$. Thus, the $R_{AHE}(H)$ component can be obtained by subtracting the $R_H \mu_0 H$ component from the $R_{xy}(H)$ data (more details in Supplementary Note 8).

**Polar MOKE and longitudinal MOKE imaging measurements.** Polar MOKE measurements were performed on a homemade optical setup. A 632.8 nm He/Ne laser beam was linearly polarized and focused on the sample surface at normal incidence by an aspheric lens with a 3-μm-diameter laser spot. The F5GT flakes were placed inside a cryostat and magnets (Montana Magneto-Optic) with an out-of-plane magnetic field up to 0.7 T and temperature down to 4 K. Upon reflection off the F5GT sample surface, the polarization axis would be rotated by the Kerr effect. The reflected beam was then directed through a Soleil-Babinet compensator to remove ellipticity before entering a Wollaston prism and split toward a pair of bridged detectors. A lock-in amplifier was used to measure the intensity variations between the separated s- and p-polarized beams, which is proportional to the Kerr rotation.

Longitudinal MOKE imaging measurements were performed with a wide-field Kerr microscope (Evico Magnetics) with an in-plane magnetic field up to 1 T and a liquid flow cryostat (300 K–4 K). In the longitudinal mode, the magnetic field is applied in-plane and parallel to the incidence plane of light, and only the in-plane component of magnetization can be obtained. As the polarized light reflects on the magnetized F5GT surface at oblique incidence, the polarization plane of reflective light rotates with a certain Kerr angle due to the Kerr effect[44,45] (details in Supplementary Fig. 7). Since different Kerr angles would be observed because of the variation in magnetization in different domains, the image contrast can be generated through an analyser. Thus, the generated image contrast can directly reflect the contrast information of magnetic domains. An Olympus 50× (LMPlanFL N) objective was used for imaging, and the spatial resolution was approximately 800 nm. The wide-field images, which contain the contrast intensity of each resolution spot of the sample surface, were recorded simultaneously while sweeping the in-plane magnetic field. The MOKE loops can also be obtained from an integration analysis of the domain contrast within the whole area of the F5GT flake.

**Magnetic anisotropy energy analysis based on the Stoner–Wohlfarth model.** Normally, in an FM system with a rhombohedral structure such as F5GT, the free energy $F$ of its magnetic moment can be described with the Stoner–Wohlfarth model considering the orientation of the magnetic easy axis as follows:

$$F = K_1 \sin^2 \theta_M - \mu_0 H M \cos(\theta_H - \theta_M) \quad (1)$$

where $K_1$ is the first-order constant of magnetic anisotropic energy to determine the magnetic easy axis, $\mu_0 H$ is the applied magnetic field, and $\theta_M$ and $\theta_H$ are the tilt angles for magnetization $M$ and $\mu_0 H$ (the normal direction of the sample plane is defined as 0° for the tilt angles, shown in the inset of Fig. 1E). As described in expression (1), at the equilibrium state with the minimum free energy $F$, $\theta_M$ should satisfy the following equation:

$$\frac{\partial F}{\partial \theta_M} = 0 = 2K_1 \sin\theta_M \cos\theta_M - \mu_0 H M \sin(\theta_H - \theta_M) \qquad (2)$$

We can obtain the following three important pieces of information from Eq. (2): a) The case of $K_1 = 0$ ($\theta_M = \theta_H$) corresponds to isotropic ferromagnetism, where the orientation of $M$ always aligns to $\mu_0 H$. b) The case of $K_1 < 0$ ($\theta_M > \theta_H$) corresponds to anisotropic ferromagnetism with an in-plane magnetic easy axis, where $M$ prefers to align in the sample plane. c) The case of $K_1 > 0$ ($\theta_M < \theta_H$) corresponds to anisotropic ferromagnetism with an out-of-plane magnetic easy axis, where $M$ prefers to align along the normal direction to the sample plane. We experimentally estimate the first-order constant $K_1$ for magnetic anisotropy energy by measuring the angular-dependent AHE (refs. [28,46]). Since we assume that $R_{AHE} \sim M_\perp = M\cos\theta_M$, $\theta_M$ can be estimated through the angle $\theta_H$ and the saturated anomalous Hall resistance $R_{AHE}^s$ based on the following expression:

$$\cos\theta_M = \frac{R_{AHE}^s(\theta_H)}{R_{AHE}^s(\theta_H = 0°)} \qquad (3)$$

where $R_{AHE}^s = R_{AHE}|_{\mu_0 H = \mu_0 H(\text{saturated})}$. We calculate $\theta_M$ as a function of $\theta_H$ using the formula:

$$\theta_M(\theta_H) = \arccos\left(\frac{R_{AHE}^s(\theta_H)}{R_{AHE}^s(\theta_H = 0°)}\right) \qquad (4)$$

The relationship between $\theta_H$ and $\theta_M$ reflects the direction of the magnetic easy axis, where $\theta_M > \theta_H$ means that the magnetization tends in the in-plane direction.

The $K_1$ value represents the total magnetic anisotropy energy density contributed from magnetic crystalline anisotropic energy $K_c$, demagnetization energy $K_{\text{demag}} = -D\mu_0 M^2/2$ (where $D$ is the demagnetizing factor), influence of strain $K_{\text{strain}}$ and surfaces $K_s$ and can be written as:

$$K_1 = K_c + K_{\text{demag}} + K_{\text{strain}} + \frac{K_s}{d} \tag{5}$$

in which the demagnetizing factor $D \approx 0$ and $D \approx 1$ for in-plane and out-of-plane magnetization orientations, and $d$ is the sample thickness.

To precisely evaluate the demagnetization energy, a correction to the magnitude ($\mu_0 H_{\text{eff}}$) and the direction ($\theta_{\text{eff}}$) of the effective magnetic field in the Stoner–Wohlfarth analysis with applied magnetic field $\mu_0 H$ should be taken into account because the local magnetic moment can generate a demagnetizing field against the external magnetic field. The correction has the following forms:

$$H_{\text{eff}} = \sqrt{(H \sin \theta_H)^2 + (H \cos \theta_H - M \cos \theta_M)^2} \tag{6}$$

$$\theta_{\text{eff}} = \arctan\left(\frac{H \sin \theta_H}{H \cos \theta_H - M \cos \theta_M}\right) \tag{7}$$

Following Stoner–Wohlfarth analysis, by inputting the value of the saturated magnetization $M$ per unit volume for F5GT ($M = 708$ emu/cm$^3$ from ref. 20) into equations (1)–(4), (6) and (7), we can obtain the values of the corrected magnetic anisotropy energy $K_{1cor}$, and the contribution of the demagnetization energy can be evaluated as $K_{\text{demag}} = K_1 - K_{1cor}$.

To precisely evaluate the contribution of surface magnetic anisotropy energy $K_s$, thickness-dependent magnetic anisotropy energy $K_{1cor}d$ is found to satisfy the following linear relation with sample thickness $d$:

$$K_{1cor}d = K_c d + K_s \tag{8}$$

where the surface magnetic anisotropy energy $K_s$ can be determined as the intercept to the vertical axis in the $K_{1cor}d$–$d$ line, and the magnetic crystalline anisotropy $K_c$ corresponds to the slope (more details in Supplementary Note 15).

In the above discussion, we use the first-order constant $K_1$ to successfully describe the magnetic anisotropy energy and the resulting magnetization direction. However, the $K_1$ value can only determine whether the magnetic easy axis is in-plane or out-of-plane. To precisely determine the canted angle of the magnetic easy axis, the free energy in the Stoner–Wohlfarth model can be extended to have both constants $K_1$ and $K_2$, as shown below:

$$F = K_1 \sin^2 \theta_\mathrm{M} + K_2 \sin^4 \theta_\mathrm{M} - \mu_0 HM \cos(\theta_\mathrm{H} - \theta_\mathrm{M}) \tag{9}$$

where $K_1$ and $K_2$ are the first- and second-order constants for magnetic anisotropic energy. Within this model, the canted angle of the magnetic easy axis can be described in a specific way, as shown in Supplementary Figs. 5, 6. Due to the ignorable planar anisotropy of the trigonal crystal structure in F5GT, it is reasonable for us to consider F5GT as an FM material with a uniaxial easy axis. Therefore, the free energy in the Stoner–Wohlfarth model does not need to be extended to a higher order to describe the planar magnetic anisotropy.

## Data availability

The data that support the plots within this paper and other findings of this study are available from the corresponding authors upon reasonable request.

## Acknowledgements

This work was supported by the A3 Foresight Program – Emerging Materials Innovation. The authors acknowledge the National Natural Science Foundation of China (91750101, 21733001, 52072168, 51861145201, 51732010), the Joint Funds of the National Natural Science Foundations of China (U21A2086), the National Key R&D Program of China (2018YFA0306200, 2021YFA1202901), and Kakenhi grant JP19H05602 from Japan Society for the Promotion of Science (JSPS).

## Author contributions

M.T., J.W.H. and F.Q. contributed equally to this work. H.T.Y., Y.I. and L.W. conceived this project. K.Z., L.W., Z.Y.L. and Y.J.T. grew F5GT crystals. A.M.N. and Y.J.T. performed STEM structural characterization. M.T., C.R.Z., L.Z. and C.Y.Q. fabricated devices. M.T., J.W.H., P.C. and Z.Y. Li performed electrical measurements. X.Y.B. took AFM thickness characterization. M.T., F.H.M. and L.L.W. performed polar MOKE measurements. M.T., F.Q., T.I., P.Z.T., H.J.Z. and X.G.W. and Y.I. analysed transport data. M.T., F.Q. and H.T.Y. wrote the manuscript with input from all authors.

## Competing interests

The authors declare no competing interests.

## Additional information

**Supplementary information** The online version contains supplementary material available at.

**Correspondence and requests for materials** should be addressed to H.T.Y., T.I. or L.W.

**Peer review information**

**Reprints and permissions information** is available at www.nature.com/reprints.

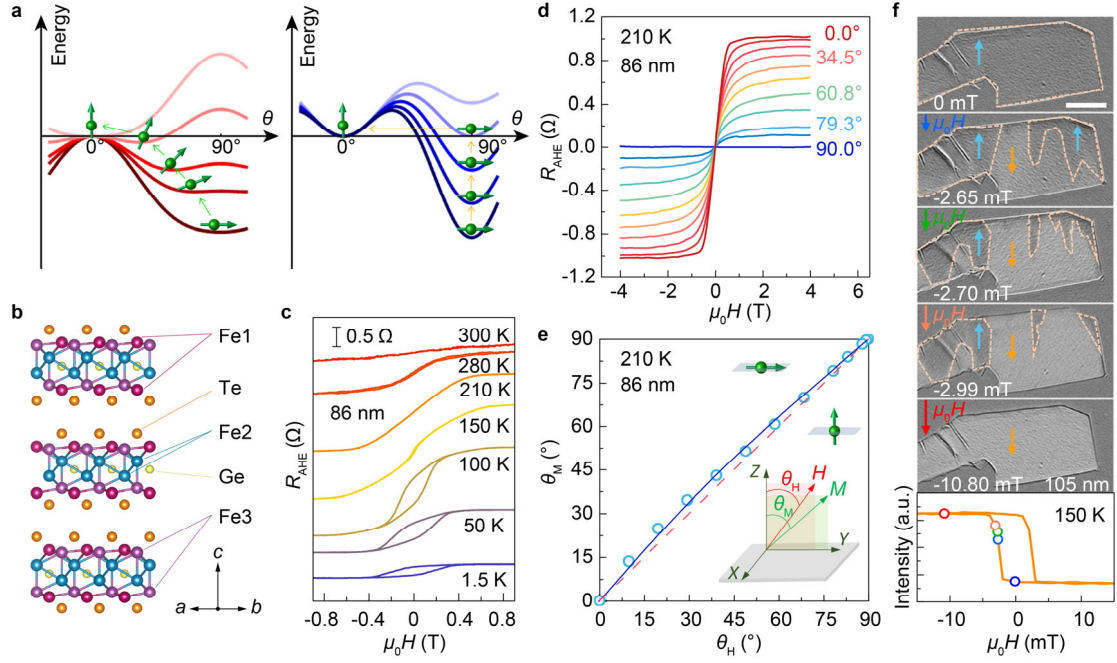

**Figure 1 Schematic concept of spin-flop and spin-flip transitions, anomalous Hall effect and magnetic domains in layered FM F5GT**

**a**, Schematic diagram of two types of magnetic transition pathways depending on the local free energy minima in FM materials. $\theta$ is the angle between the magnetic easy axis and the normal direction of the sample surface. Green arrows with balls represent the direction of magnetic easy axes. **b**, Crystal structure of F5GT viewed from the [110] direction. The atom sites are marked as Fe1 (magenta balls), Fe2 (blue balls), Fe3 (purple balls), Ge (yellow balls) and Te (orange balls). Fe1 atoms occupy split sites above or below the Ge site (for more details, see Supplementary Note 1). **c**, $R_{\text{AHE}}$ as a function of the magnetic field at different temperatures for a thick flake (86 nm). **d**, $R_{\text{AHE}}$ as a function of the magnetic field of this sample at 210 K with varying $\theta_H$ from 0° to 90°. $\theta_H$ is defined as the angle between the direction of the applied magnetic field and the normal direction of the sample surface. The $R_{\text{AHE}}(H)$ data are obtained from the $R_{xy}(H)$ data. **e**, Relation between $\theta_M$ and $\theta_H$ at 210 K obtained from **d**, plotted with blue hollow circles. $\theta_M$ is the angle between the direction of magnetization $M$ and the normal direction of the sample surface. The blue line is the fitting curve based on the Stoner–Wohlfarth model. The dashed line represents the relation of $\theta_M = \theta_H$, indicating $K_1 = 0$. The upper-left zone represents $\theta_M > \theta_H$ with $K_1 < 0$, indicating the in-plane magnetic easy axis, while the bottom-right zone represents $\theta_M < \theta_H$ with $K_1 > 0$, indicating the out-of-plane magnetic easy axis. The inset shows the schematic measurement geometry for the angle-dependent AHE. **f**, Magnetic domain images and MOKE loop of an F5GT thick flake obtained in longitudinal mode with a wide-field Kerr microscope at 150 K. The dark and bright regions of the F5GT flake correspond to the magnetic domains with opposite directions of magnetization, indicated by blue and orange arrows, respectively. The yellow dashed lines highlight magnetic domain boundaries. The applied magnetic field is applied in-plane (marked by upper-left arrows in each image) and is parallel to the incidence plane of the light. The bottom panel shows the MOKE loop in which the intensity is integrated by the whole flake region while sweeping the in-plane magnetic field. The circles on the MOKE loop correspond to the magnetic field at which the magnetic domain images shown above were taken. The scale bar is 20 μm.

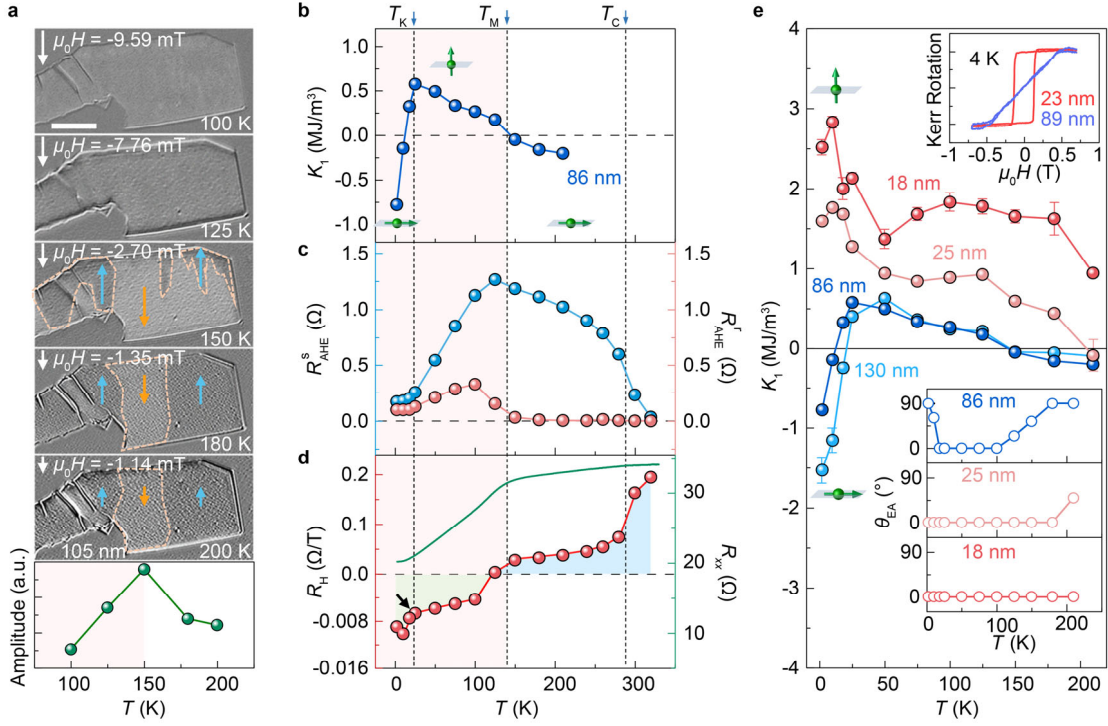

**Figure 2 Temperature- and thickness-dependent magnetic anisotropy in F5GT devices**

**a**, Magnetic domain images at different temperatures obtained in longitudinal mode with a wide-field Kerr microscope setup. The magnetic field is applied in-plane (labelled by white arrows) and is parallel to the incidence plane of the light. The dark and bright regions of the F5GT flake in the magnetic domain images correspond to the opposite directions of magnetization, indicated by blue and orange arrows, respectively. The bottom panel shows the temperature evolution of the amplitude of the MOKE loop in which the intensity is integrated by the whole flake region during sweeping of the in-plane magnetic field. The scale bar is 20 μm. **b**, Temperature-dependent $K_1$ values of the flake with a thickness of 86 nm. The blue arrows and the corresponding dashed lines indicate the three magnetic transition temperatures at $T_C$, $T_M$ and $T_K$, respectively. **c**, Temperature dependence of saturated $R_{AHE}^s$ ($R_{AHE}|_{\mu_0 H=2\,T}$, blue balls) and remnant $R_{AHE}^r$ ($R_{AHE}|_{\mu_0 H=0\,T}$, pink balls). **d**, Hall coefficient $R_H$ (red balls and line) and longitudinal resistance $R_{xx}$ (green line) as a function of temperature. The positive and negative parts of the $R_H$ axes are plotted with different linear scales. The black arrow indicates the temperature at $T_K$. $R_H$ is deduced from the $R_{xy}(H)$ data. **e**,

Temperature-dependent $K_1$ for F5GT flakes with different thicknesses: 130 nm, 86 nm, 25 nm, and 18 nm. Upper inset: Kerr rotation for a thick flake (89 nm) and a thin flake (23 nm) acquired by polar MOKE measurements at 4 K. Bottom inset: temperature-dependent $\theta_{EA}$ for flakes with thicknesses of 86 nm, 25 nm, and 18 nm. Green arrows together with a plane in **b** and **e** depict in-plane and out-of-plane magnetic easy axes.

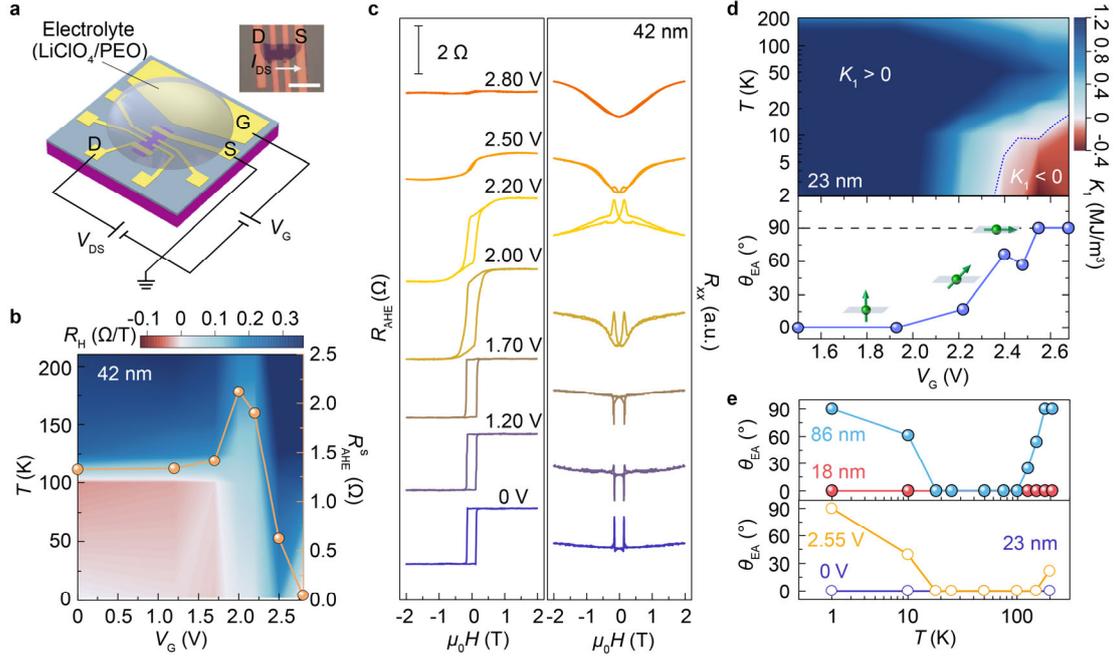

**Figure 3 Electrical tuning of magnetic states and MAE in gated F5GT devices**

**a**, Schematic illustration and optical image of the Hall-bar geometry for electrolyte gating. A solution of LiClO$_4$ and PEO was used as the gating electrolyte. The letters D, S, and G represent the electrodes of the drain, source, and gate for electric measurements, respectively. $I_{DS}$, $V_{DS}$, and $V_G$ indicate the current and voltage between D and S and the gate voltage, respectively. The scale bar is 20 μm. **b**, Color mapping of $R_H$ as a function of temperature and $V_G$ and the $R_{AHE}^s$–$V_G$ data at 2 K (orange balls and line). **c**, Hysteresis loops of $R_{AHE}$ and magnetoresistance at 2 K with varying $V_G$. $R_{xx}$ is shown in arbitrary units. **d**, Top panel: color mapping for the temperature-dependent $K_1$ of a device with a thickness of 23 nm under different gate voltages. The blue dotted line highlights the boundary where $K_1 = 0$. Bottom panel: $\theta_{EA}$ as a function of $V_G$ at 2 K. Green arrows together with a plane depict the direction of magnetic easy axes. **e**, Top panel: temperature-dependent $\theta_{EA}$ of the device with thicknesses of 86 nm and 18 nm with no gate voltage. Bottom panel: temperature-dependent $\theta_{EA}$ of the device with a thickness of 23 nm under two different $V_G$ at 2 K.

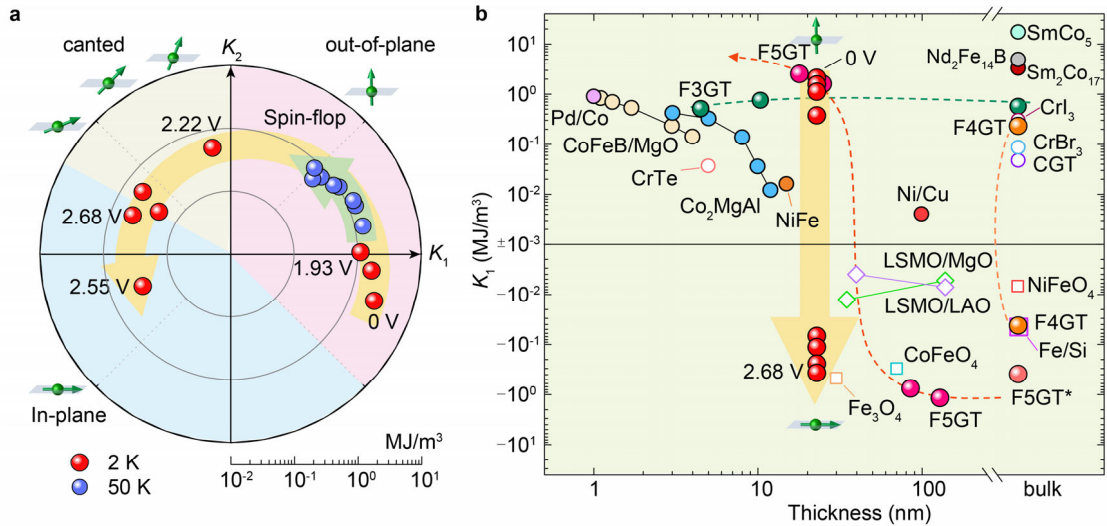

**Figure 4 Gate-tuned spin-flop transition in F5GT devices**

**a**, Polar chart of gate-dependent $K_1$ and $K_2$ of the flake with a thickness of 23 nm at 2 K (red balls) and 50 K (blue balls). Different colored areas represent three different spin orientation states: in-plane (light blue), out-of-plane (light pink), and canted spin orientations (light yellow). The $K_1$ and $K_2$ values at $V_G$ = 0, 1.50, 1.93, 2.22, 2.40, 2.48, 2.55, and 2.68 V illustrates that the magnetic easy axis can be continuously tuned from the out-of-plane magnetization zone, across the canted magnetization zone, and toward the in-plane magnetization zone. The evolution of data points for 2 K and 50 K are guided by the yellow arrow and green arrow. Green arrows together with a plane depict the direction of magnetic easy axes. **b**, Survey of the $K_1$ values of MAE for classic FM materials. The black horizontal line labelled at $\pm 10^{-3}$ MJ/m³ denotes the boundary between positive and negative $K_1$, ignoring the region with an absolute value of $K_1$ smaller than $10^{-3}$ MJ/m³. The dashed lines show thickness-dependent $K_1$ in Fe$_n$GeTe$_2$ ($n$ = 3,4,5), guiding the eyes. The gate-dependent $K_1$ of the 23-nm-thick flake (red balls) evolves from positive to negative with increased $V_G$, highlighted by the yellow arrow. Green arrows together with a plane depict the direction of magnetic easy axes. The specific values of the data points are listed in Supplementary Tables 2–4.